# Visible-Light-Gated Reconfigurable Rotation of Nanomotors in Electric Fields

Zexi Liang[†], Donglei (Emma) Fan[†,‡]

[†]Materials Science and Engineering Program, The University of Texas at Austin, Austin, TX, 78712

‡ Department of Mechanical Engineering, The University of Texas at Austin, Austin, TX, 78712

**Abstract:** Highly efficient and widely applicable working mechanisms that allow nanomaterials and devices to respond to external stimuli with controlled mechanical motions could make far-reaching impact to reconfigurable, adaptive, and robotic nanodevices. Here, we report an innovative mechanism that allows multifold reconfiguration of mechanical rotation of semiconductor nanoentities in electric ($E$) fields by visible light stimulation. When illuminated by light in the visible to infrared range, the rotation speed of semiconductor Si nanowires in electric fields can instantly increase, decrease, and even reverse the orientation depending on the intensity of the applied light and the AC $E$-field frequency. This multifold rotation configuration is highly efficient, instant, and facile. Switching between different modes can be simply controlled by the light intensity at an AC frequency. An array of experimentations, theoretical analysis, and simulations are carried out to understand the underlying principle, which can be attributed to the optically tunable polarization of Si nanowires in aqueous suspension and an external electric field. Finally, leveraging this newly discovered effect, we successfully differentiate semiconductor and metallic nanoentities in a non-contact and non-destructive manner. This research could inspire a new class of reconfigurable nanoelectromechanical and nanorobotic devices for optical sensing, communication, molecule release, detection, nanoparticle separation, and microfluidic automation.

Changing the working scheme of nanodevices from static to dynamic, from passive to active, enabling intelligent and autonomous performances, could bring unprecedented impact to an array of applications in electronics, communication, sensing, therapy, and single-cell biology research[1-16]. Mechanically active materials and structures, which change volume, shape, and mechanical motions in response to external stimuli, are essential for realizing intelligent and autonomous electronics and have received immense research interest [17-22]. For instance, the widely used shape memory alloys that undergo controlled volume change in external physical fields are now key elements of actuators widely used in aerospace, automobile, and precision instrumentation.[23] Miniaturized mechanical grippers, made of strategically fabricated multilayer thin films with tailorable interfacial stresses, can readily self-fold to capture live cells and sample biologic tissues when sensing the acidic environment near the cells [24, 25]. More recently, mimicking live microorganisms in nature, artificial micro/nanorobotic devices have been fabricated with functionality analogous to the behavior of natural organisms.[26, 27] Acoustic hologram technique assembles particles into reconfigurable arbitrary patterns in time evolving acoustic fields.[28] Many techniques rely on light for motion control. Azobenzene chemical groups form shapes reconfigurable with light-induced isomerization in both macroscopic and molecular scales. These polymers are applied to carry out complex tasks, such as the capture and release of DNA molecules [29-31]. Self-electrophoresis, based on photochemistry, is exploited to develop light-guided micromotors with intrinsic asymmetric structures or light-induced asymmetry,[32] and a transition of $TiO_2$ microspheres is observed from swarm-like aggregation to mono-dispersive suspension on exposure to UV-light.[33] The redistribution of *E*-fields in response to light patterns on photoconductive substrates can readily translocate nanoparticles, cells or droplets by dielectrophoresis or electrowetting [34, 35]. Plasmonic structures make it possible to control light in subwavelength scope. With uniquely designed metallic structures, torques can be induced from laser and compel the rotation. [36, 37] Thus, these remarkable materials that respond to external stimuli with mechanical actuations are highly potential for many unprecedented applications. However, most such materials to date only respond to stimuli in one manner, *e.g.* linear translocation, bending, rotation, or aggregation, and just work in limited environments. Moreover, to obtain these materials, sophisticated design, synthesis, and functionalization are often required. To overcome these bottlenecks, it is highly desirable to investigate a completely new mechanism in activating nanomaterials with efficient, reversible, and multifold reconfigurabilities.

In this work, by controlling light stimuli, we achieved multifold mechanical reconfigurability of nanomechanical devices, *i.e.* rotary nanomotors, in an efficient, versatile, and potentially widely applicable manner. The investigation is based on exploitation of the underlying interactions of electrokinetics around semiconductor Si nanowires, optically controlled interfacial polarization of Si nanowires in aqueous suspension and the applied high-frequency electric fields. Reconfigurable mechanical rotation of nanowires, including acceleration, deceleration, stop, and reversal of orientation are all realized by controlling illumination of visible light. The working mechanism is understood by both theoretical modeling and numerical simulations. With this newly discovered opto-mechanical effect, we demonstrate efficient non-contact differentiation of mixed metallic (Au) and semiconductor (Si) nanowires. This work could open new opportunities in identification and separation of nanoparticles and live entities and inspires approaches to equip nanomaterials and devices with multifold light-controlled mechanical reconfigurability for the emerging nanorobotics.

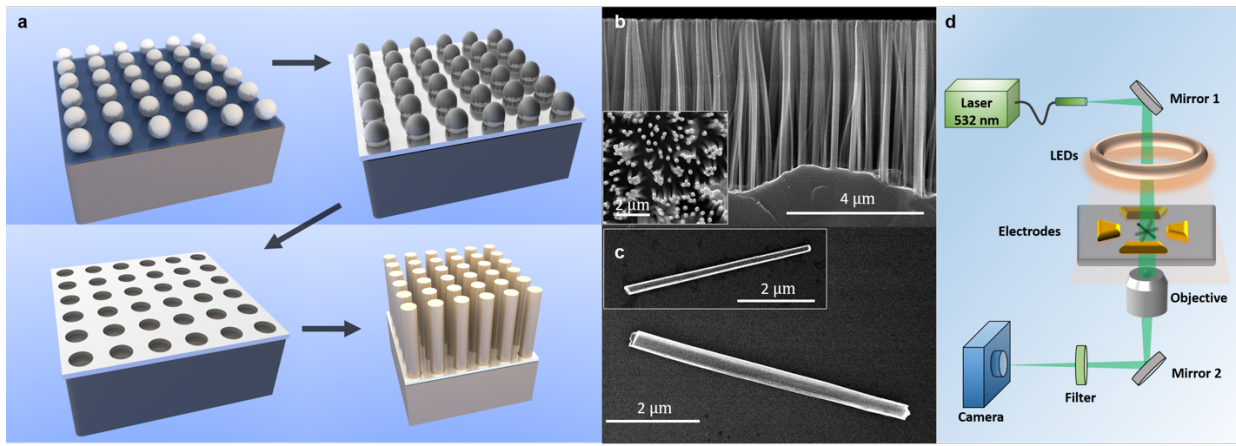

Figure 1. Silicon nanowire fabrication process and experimental setup. (a) Process of silicon nanowire fabrication via nanosphere lithography masked metal-assisted-chemical-etching. (b-c) Scanning Electron Microscopy images of silicon nanowires. (b) Cross-section view of arrays of silicon nanowires of 200 nm diameter made from n-type 560-840 $\Omega \cdot cm$ silicon wafer. (inset: top-view) (c) A nanowire of 500 nm in diameter after sonication from the wafer. (inset: a nanowire of 200 nm in diameter) (d) Schematic of the optical-electric system setup for reconfigurable manipulation.

## Results

**Fabrication of silicon nanowires.** Arrays of silicon nanowires (diameter: 100 nm and above) are fabricated by modified metal-assisted chemical etching templated through nanosphere assemblies [38-40]. The process is shown in Fig. 1A and discussed in detail in the methods part. In brief, a monolayer of nanospheres is assembled on the silicon wafer as masks for the Ag film deposition, which is then removed, leaving the Ag film with arrays of nanoholes. By immersing the sample in the etchant of hydrofluoric acid and hydrogen peroxide, Si nanowires are fabricated with controlled diameters as shown in the SEM images in Fig. 1(b) and (c). For nanowires with diameters less than 100 nm, one-step metal-assisted chemical etching is instead adopted.[41] A droplet of nanowire suspension (20 μl) is dispersed at the center of a quadruple microelectrode (500 μm in gap) made of Au (100 nm)/Cr (5 nm) on glass for testing. After allowing several minutes for the nanowires deposit to the bottom, their motions are recorded and analyzed. If not stated otherwise, the nanowires are 500 nm in diameter and 5 μm in length, fabricated from n-type 560-840 $\Omega \cdot cm$ silicon wafers.

**Light-controlled reconfigurable electrorotation.** The microelectrode is placed on an inverted microscope equipped with a 532 nm as the stimulating source and a customized light-emitting-diode ring with low illuminance (500 lux) as the background (Fig. 1d). By applying four AC voltages, each phase-shifted 90 degrees from the last, of frequencies ranging from 2.5 kHz to 2 MHz on the quadruple electrodes, a rotating AC electric field can be readily created to rotate the nanowires, an effect termed electrorotation. The distinct rotation characterizations with and without laser illumination versus AC frequencies from 5 kHz to 2 MHz at 20 Vpp is shown in Fig. 2a (Movie S1-4).

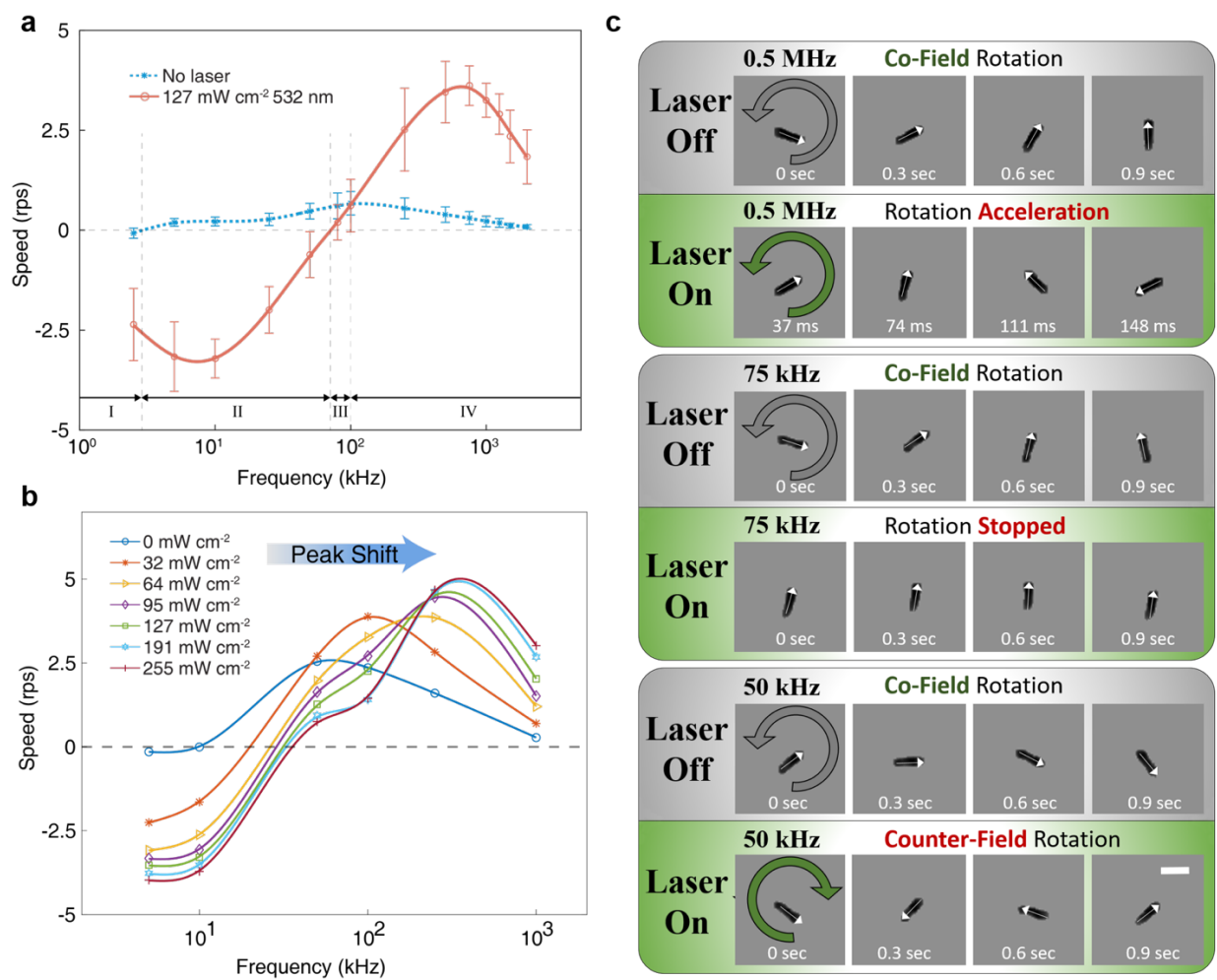

Figure 2. Rotation of silicon nanowires and the reconfiguration by laser illuminance in electric fields. (a) Rotation speed versus AC frequency of silicon nanowires (n-type 560-840 $\Omega$ cm, 5 $\mu$m length and 500 nm diameter) in dim environment and under 532 nm laser illumination (127 $mW\ cm^{-2}$). (b) The rotation versus AC frequency of silicon nanowires illuminated at different laser intensities. (20 Vpp at all frequencies unless stated otherwise. The trend lines are obtained with cubic spline interpolation.) (c) Snapshots of rotation of single silicon nanowire at AC frequencies 50 kHz-0.5 MHz, showing three distinct optical responses. Scale bar, 5 $\mu$m.

In the following, we name the blue and red curves with and without laser stimuli as the bright and dim spectrum, respectively. The positive and negative signs of the rotation indicate the co-field and counter-field rotation of the nanowires to that of the external electric fields, respectively. The dim spectrum in Fig. 2a, obtained at low light intensity (~500 Lux), shows consistent co-field rotation in AC electric fields from 5 kHz to 2MHz with a peak near 100 kHz. At low frequencies, *e.g.* 2.5 kHz, some nanowires in suspension exhibit counter-field rotation, different from the rotational spectra reported previously [40], which could be attributed to the differences in size, materials, and light conditions. However, under high-intensity illumination, the reversal of the rotation orientation is clearly observed at frequencies less than 50 kHz as shown in the bright spectrum (red curve) in Fig. 2a. Besides the reversal of the rotation orientation, laser illumination substantially increases the absolute rotation speeds of the nanowires across the entire frequency region except around the transition frequency at ~80 kHz. Two rotation peaks can be found at 10 kHz and 750 kHz, corresponding to the maximum counter-field and co-field rotation, respectively. Overall, the effect of laser to the rotational spectra can be divided into four frequency regions: Region I: low AC frequencies, below 5 kHz, increased rotation speed in the counter-field direction; Region II: from 5kHz to 80 kHz, reversed rotation orientation from co-field to counter-field; Region III: from 80 kHz to 100 kHz, lowered rotation speed in the original co-field direction; Region IV: from 100 kHz to 2 MHz, increased rotation speed in the original co-field direction.

After discovering the dramatic reconfigurable rotation of Si nanowires with optical illumination, we systematically investigated the effect of laser intensity on rotational spectra by tracking a single nanowire through all intensity conditions and frequencies (Fig. 2b). Obviously, the increase of the laser power from 0 to $255 \text{ mW cm}^{-2}$ increases the rotation speed at both high and low frequencies, while the peak positions of the co-field rotation blue shift from ~50 kHz to above 300 kHz monotonically.

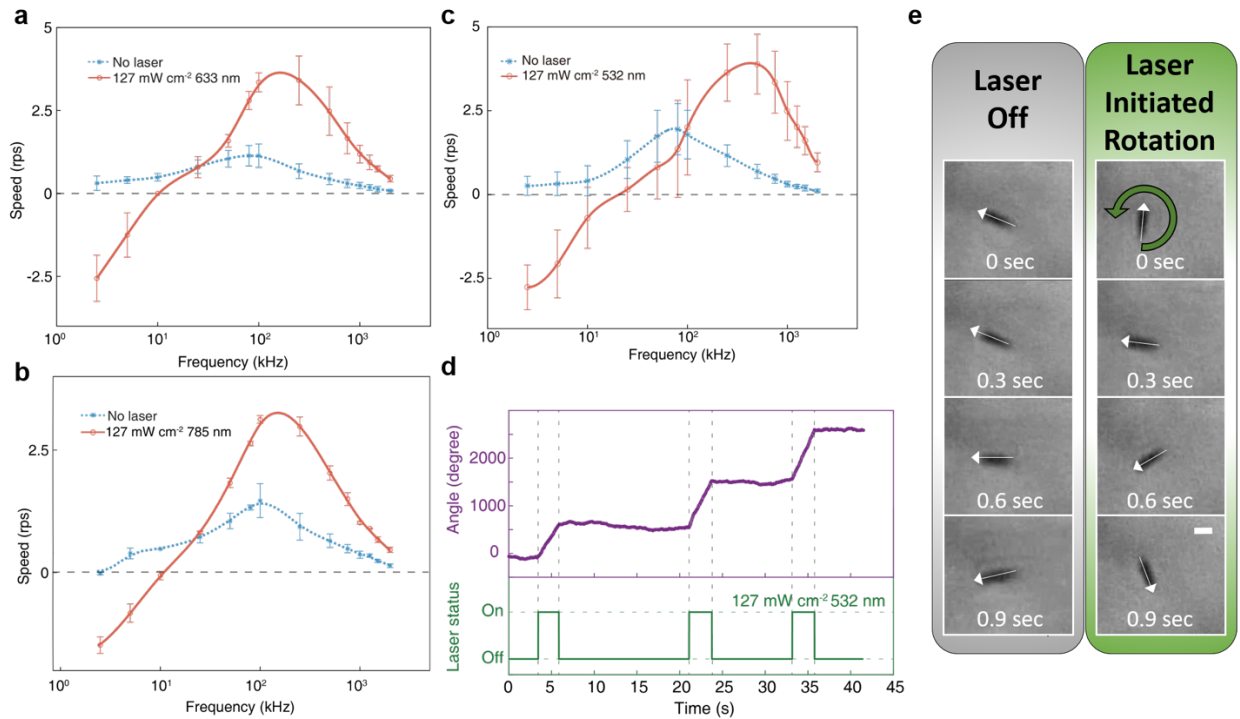

Figure 3. Reconfigurable rotation at different laser wavelengths and size of nanowires in an AC electric field. (a-b) Rotational spectra of silicon nanowire under illumination of 633 nm and 785 nm laser, same

nanowires used as in Fig. 2a. (c) Rotational spectra of silicon nanowire of 200 nm in diameter (n-type 560-840 $\Omega$ cm). (d) The rotation angle versus time showing optical initiated rotation of nanowires of sub 100 nm in diameter. (0.5 MHz, 20 Vpp) (e) Snapshots of laser induced rotation of a nanowire of sub 100 nm in diameter. Scale bar 2 $\mu$m.

A similar opto-responsive rotation is observed with laser illumination at 633 nm and 785 nm (127 mW cm$^{-2}$) as shown in Fig. 3a, b respectively. How the mechanical rotation of nanowires responding to light of different wavelengths is determined by their material properties as well as the geometries, which will be further investigated in a separation effort. In this work, we found the versatile reconfiguration of Si nanowires is widely attainable with a light source from the visible to near infrared regime, which is desirable for a variety of applications.

The above strong light sensitivity in electro-rotation is obtained from Silicon nanowires of 500 nm in diameter. It is interesting to know if such an optical reconfigurable effect can be obtained from nanowires with smaller diameters, *i.e.* less than 100 nm. To determine this, we fabricated a series of nanowires with of 200 nm and <100 nm in diameters and tested their optical responses. Indeed, the typical reconfiguration behaviors observed in 500 nm Si nanowires are obtained in nanowires of 200 nm diameter as shown in Fig. 3c. The method still holds for smaller nanowires less than 100 nm in diameter. Although their movement due to Brownian motion is relatively stronger, we still observed the dramatic change of rotation speed when turning the laser on and off (Fig. 3d). This is also shown in the snap shots in Fig. 3e (Movie S5). When the 532 nm laser (127 mW cm$^{-2}$) is on, the nanowire instantly switches to continuous rotation from random Brownian movement at an AC field of 0.5 MHz.

These experimental studies demonstrate that the manner of reconfiguration of the nanowires, including acceleration, deceleration and reversal of rotation chirality, depends on factors including AC frequency, laser power, and dimensions of silicon nanowires. The optical response of electrorotation is applicable to Si nanowires with wide range of sizes as well as various resistivity from intrinsic (>5000 $\Omega$ cm) to moderately n-doped silicon (0.6-0.75 $\Omega$ cm), while the highly n-doped degenerate silicon nanowire (0.001-0.005 $\Omega$ cm) does not response to the light, which will be discussed later. The same effect could also be observed in other types of photosensitive semiconductors and polymers. To understand these complex phenomena, we conducted both theoretical analysis and numerical simulations.

**Theoretical modelling and analysis.** The rotational torques on small particles arise from the interactions of polarized particles and high frequency rotating *E*-field.[42] The electric polarization of a nanoparticle is governed by multiple parameters including permittivity, conductivity and geometry of the nanoparticle and the electric properties of suspension medium, as well as frequency of the rotating *E*-field. Theoretical models have been developed to understand the rotational behavior of metal and semiconductor nanowires. In additional to the long established Maxwell-Wagner interfacial polarization effect,[42] the electrical double layer (EDL) charging effect is identified recently that also governs the rotation. [43, 44]

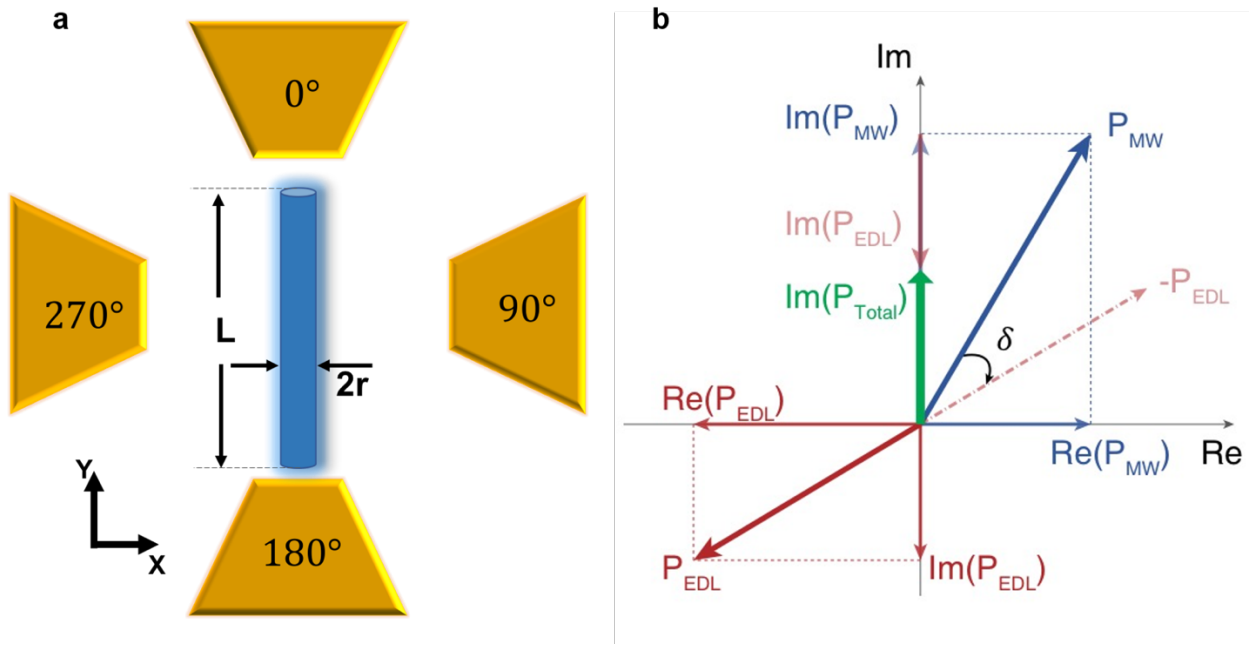

Figure 4. Modeling of nanowire polarization. (a) Model of a Silicon nanowire as a cylinder with length *L* and radius *r*. (b) Schematic of the real and imaginary dipole moment components induced by the Maxwell-Wagner relaxation and electrical double layer effect.

When a rotating AC *E*-field is applied via the quadruple electrodes, an electric torque is exerted on the induced dipole moment of the nanowire and compels its rotation. Consider a nanowire placed in the center of the quadruple electrodes with the long axis pointing along the y direction (Fig. 4a, top-down view). The nanowire is modeled as a long cylinder with a radius of $r$ and a length of L. A rotating AC *E*-field is generated by applying four AC voltages with $90°$ sequential phase shift on the quadruple electrodes. The *E*-field can be given by $\mathbf{E}(t) = E_0 \text{Re}[(\hat{\mathbf{x}} - i\hat{\mathbf{y}})e^{i\omega t}]$, for the counterclockwise rotation. The nanowire is polarized with a total dipole moment ($\underline{\mathbf{p}}$), expressed as $\underline{\mathbf{p}} = \underline{\alpha_x}\, \underline{\mathbf{E_x}} + \underline{\alpha_y}\, \underline{\mathbf{E_y}} = \underline{\mathbf{p_x}} + \underline{\mathbf{p_y}}$, where $\underline{\alpha_x}$ and $\underline{\alpha_y}$ are the complex polarizability along the transverse and longitudinal axes of the nanowire, respectively. The underbars denote the complex variables with phasor. Since the angular frequency of the AC *E*-field is much higher than that of the rotation speed of the nanowire, in the frequency region that we are studying, we can assume that the nanowire remains still while the AC *E*-field oscillating one cycle, and then the time-averaged electrical torque can be readily written as:

$$\tau_e = \frac{1}{2}\mathbf{Re}\left[\underline{\mathbf{p}}(t) \times \underline{\mathbf{E}}^*(t)\right] = -\frac{1}{2}E_0^2 \mathbf{Im}[\underline{\alpha_x} + \underline{\alpha_y}]\hat{z}. \tag{1}$$

Both the Maxwell-Wagner relaxation process and the effect of EDL contribute to the complex electric polarizability of the nanowire in an aqueous suspension. For a silicon nanowire suspended in DI water in an *E*-field, charges accumulate at the interface between the two dielectrics due to the differences in electrical properties, including the permittivity and electric conductivity. As the polarization involves several relaxation processes, the amount of charges accumulated at the interface depends on the AC frequencies, and thus, resulting in a frequency dependent induced dipole moment, which is referred as the Maxwell-Wagner polarization or relaxation. Meanwhile, once the interfacial polarization is established, the ions in suspension will be attracted towards the particle surface to screen the surface charge, which forms into an EDL. In general, we consider the polarization as a two-step process, in which the external *E*-field first induces the interfacial polarization and charges the surface, then followed by the formation of EDL due to the surface charge attraction. Since the relaxation processes of both Maxwell-Wagner polarization and EDL take finite time, there exists a frequency dependent phase lag between the Maxwell-Wagner polarization and the external *E*-field, as well as a frequency dependent phase lag between the Maxwell-Wagner polarization and EDL under an AC *E*-field. Due to the synergistic effect, the rotational spectrum of Si nanowires is obtained as observed in Fig. 4b.

The modeling of the system is carried out analytically. The Clausius-Mossotti factor ($K_i$) is the key term governing the Maxwell-Wagner polarizability along one of the principle axes (*i*) of a nanowire modeled as a prolate ellipsoid. It is calculated as:

$$K_i = \frac{\varepsilon_p - \varepsilon_m}{3(L_i(\varepsilon_p - \varepsilon_m) + \varepsilon_m)} \tag{2}$$

where $\varepsilon_p, \varepsilon_m$ are the complex dielectric constants of the nanowire and the medium defied as: $\varepsilon_j = \varepsilon_0 \varepsilon_{r_j} - i\frac{\sigma_j}{\omega}$ (the subscript "j" is *p* or *m*, referring to the particle or medium) and $L_i$ is the depolarization factor along one of the major axes ($a_y = L/2, a_x = a_z = r$) given by:

$$L_i = \frac{a_x a_y a_z}{2} \int_0^\infty \frac{ds}{(s+a_i^2)B}, \tag{3}$$

where $B = \sqrt{(s+a_x^2)(s+a_y^2)(s+a_z^2)}$. Therefore, the overall in-plane dipole moment due to the Maxwell-Wagner relaxation ($p_{MW}$) can be expressed as:

$$p_{MW} = 4\pi a_x a_y a_z \varepsilon_m (K_x E_x + K_y E_y). \tag{4}$$

Both the real and imaginary parts of $\alpha_x$ and $\alpha_y$ are calculated. However, since the polarization of nanowire along the longitudinal y-direction is much greater than that in the transverse x-direction, one may approximate by only focusing on the dominating $\alpha_y$.

Next, we analyzed how the electric double layer (EDL) contributes to the imaginary part of the electric polarization in Eqs. (1). We modeled the EDL by using an equivalent RC circuit with a time constant of $\tau_{RC}$ to understand the electrokinetic phenomena observed in our experiments. The EDL is formed due to the surface charge accumulation of Maxwell-Wagner polarization, instead of direct response to the external *E*-field. While in a high-frequency AC field, the charging and discharging of the EDL cannot completely follow the Maxwell-Wagner polarization of the nanoparticle, which causes a phase lag. In a low frequency limit, the charges due to the Maxwell-Wagner polarization on the nanowire can be fully screened by the EDL, resulting in $p_{EDL} \approx -p_{MW}$, where the total dipole moment essentially cancels out with a negligible negative residue (0 <

$\frac{|\mathbf{p}_{EDL}|-|\mathbf{p}_{MW}|}{|\mathbf{p}_{MW}|} \ll 1$) as the thickness of EDL is much smaller compared to the size of the nanowire. In a high AC frequency, given the RC model, the resulted magnitude of dipole moment of EDL can be expressed as $|\mathbf{p}_{EDL}|=|\mathbf{p}_{MW}|\frac{1}{\sqrt{\omega^2\tau_{RC}^2+1}}$. Here, we need to take phasor into consideration due to the phase lag between $-\mathbf{p}_{EDL}$ and $\mathbf{p}_{MW}$, the tangent of the phase lag ($\delta$) is calculated as $\tan\delta = -\omega\tau_{RC}$, the negative sign of which indicates a phase delay as shown in Fig. 4b. Therefore, the Imaginary part of EDL induced dipole moment $\text{Im}(\mathbf{p}_{EDL})$ is calculated as:

$$\text{Im}(\mathbf{p}_{EDL}) = -[\text{Re}(\mathbf{p}_{MW})\sin\delta + \text{Im}(\mathbf{p}_{MW})\cos\delta]\frac{1}{\sqrt{\omega^2\tau_{RC}^2+1}}. \qquad (5)$$

The total electric dipole moment of the nanowire system ($\mathbf{p}_{total}$) includes both the Maxwell-Wagner relaxation process ($\mathbf{p}_{MW}$) and the EDL component ($\mathbf{p}_{EDL}$). The imaginary component of $\mathbf{p}_{total}$ is given by [Fig. 4(b)]:

$$\text{Im}(\mathbf{p}_{total}) = \text{Im}(\mathbf{p}_{EDL}) + \text{Im}(\mathbf{p}_{MW}). \qquad (6)$$

The values of different parameters used in the calculation are listed in Table S1 (supporting information), in which $\tau_{RC}$ is chosen to best fit the experimental data of Fig. 2a. The electric conductivity of Si is systematically swept from $1\times10^{-3}$ S m$^{-1}$ to $1$ S m$^{-1}$.

The nanowire suspended in water with maximum rotation speed less than 5 r.p.s. is within the low Reynolds number region (Re $\ll 1$), and thus the system is in viscous laminar flow limit, free of turbulence with a constant drag coefficient of[45]:

$$\gamma = \frac{\tau_\eta}{\omega} = \frac{\pi\eta l^3}{3(\sigma-1.14-0.2\sigma^{-1}-16\sigma^{-2}+63\sigma^{-3}-62\sigma^{-4})} \qquad (7)$$

where $\sigma = \ln\frac{L}{r}$. For a nanowire with 500 nm diameter and 5 μm length, the drag coefficient is calculated as $8.3\times10^{-20}$ N m s rad$^{-1}$. Therefore, the rotation spectrum of Si nanowires versus AC frequencies can be readily calculated theoretically by combining Eqs. (1) and (4-7) as shown in Fig. 5a.

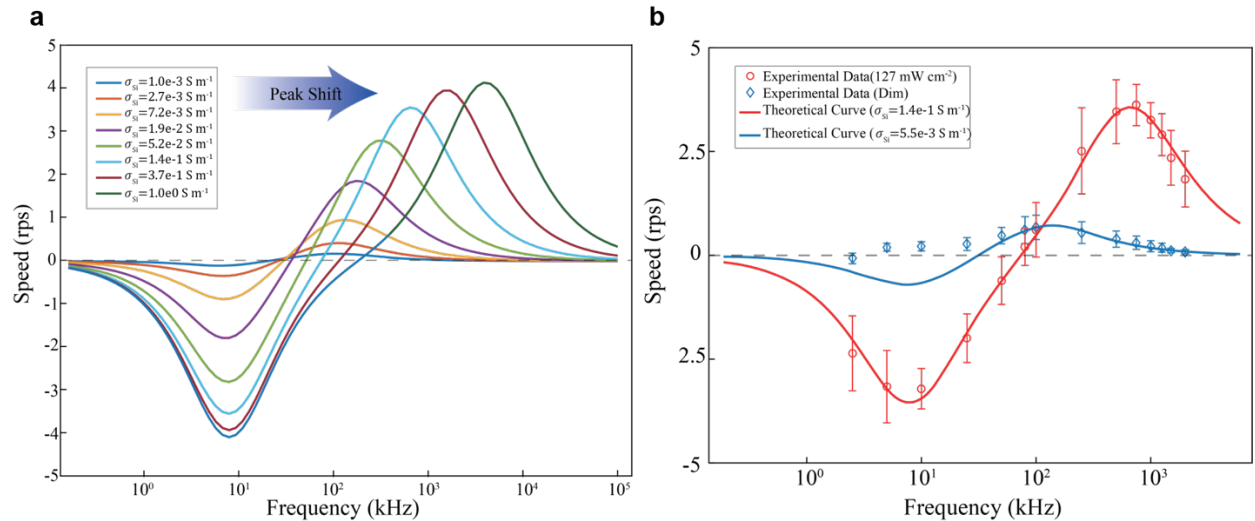

Figure 5. Theoretical calculation of rotation speed versus AC frequency of silicon nanowire (L = 5 μm and r = 250 nm). (a) Theoretical calculation of the rotation of nanowires of different electric conductivities

versus AC frequency. (b) Theory compared to the experimental results. The two particular conductivities represent the previous dim and bright conditions, respectively.

It is known that for semiconductors, photon excitation can generate additional free charge carriers, resulting in additional electric conductivity, the so-called photoconductivity. Previous studies show Si nanowires have strong photon response,[46-48] where the increase of electric conductivity can reach several orders of magnitude. When we monotonically increase the electric conductivity of Si from $1\times10^{-3}$ S m$^{-1}$ to 1 S m$^{-1}$, the peaks of both the co-field and counter-field rotations blue shift and the magnitude of both increase substantially (Fig. 5a). These spectrum characteristics obtained from the calculation agrees reasonably well with those obtained in our experiments as shown in Fig. 5b. It suggests that the enhanced electric conductivity of Si due to light illumination is the key factor that leads to the observed reconfigurability of rotation of nanowires. The theoretical fitting of the bright spectrum agrees well through the entire frequency range for an electric conductivity of $1.4\times10^{-1}$ S m$^{-1}$. Here we noticed that this conductivity is close to that of the wafer from which the nanowires are synthesized. However, as shown by previous works, due to the surface depletion effects, the carrier density of Si nanowires is usually much lower than that of the wafer, since surface states at the interface between silicon and natural oxidization layer trapped most carriers from dopants (supplementary note 3).[49-52] With numerical simulation and calculation with consideration of light absorption and recombination rate, we further estimated the photoconductivity of Si nanowires, which is on the same order of magnitude of that obtained from the fitting. It suggests the good feasibility of the modeling. It also indicates that the photon-generated carriers contribute majorly to the effective conductivity of silicon at our light illumination conditions (supplementary note 4).

The theoretical fitting to dim spectrum agrees in high frequencies while deviates in low frequencies, *e.g.* the counter-field rotation is shown in fitting but not observed in these experiments (Fig. 5b). This could be attributed to the simplified consideration of the charge distribution in nanowires that induce EDL when light is weak (dim environment). In reality, when light is dim, the carrier density is low and the Maxwell-Wagner polarization induced net charges not only accumulate at the surface, but also distribute inside the silicon nanowire. Since the EDL is only sensitive to the surface charge, ineffective in responding to charges away from the surface, a weaker EDL compared to that used in the theoretical analysis is expected, which could account for the observed difference between the calculation and experiments at low frequencies.

Similar reconfigurable behaviors are observed in a wide range of Silicon nanowires made from n-doped wafers of 0.6- 0.75 $\Omega$ cm to intrinsic Si of over 5000 $\Omega$ cm (Fig. S1). For the Si nanowires made from Si wafers of 0.001-0.005 $\Omega$ cm with a high doping level of ~$10^{19}$ cm$^{-3}$, the optical stimuli can no longer change the rotation behaviors (Fig. S1). It can be attributed to the ultrahigh dopant concentration of $1\times10^{19}$ cm$^{-3}$ of the Si wafer, which corresponds to a carrier concentration 6 orders of magnitude higher than that induced by the laser excitation ($1.8\times10^{13}$ cm$^{-3}$) as shown in the calculation in supplementary note 4. The results further agree with our understanding that the increase of carrier density from photoexcitation leads to the responsiveness of the mechanical rotation of nanowires. Overall, the balance between the generation of photoexcited carriers and recombination, as well as the surface depletion effect explain the observation of the general similarity of rotational spectra of nanowires made from silicon wafers with resistivity from 0.6 $\Omega$ cm to over 5000 $\Omega$ cm, as well as the vanishing of the optical response in highly doped silicon nanowires with a resistivity of 0.001-0.005 $\Omega$ cm.

In the above analysis, we carefully omitted the light induced heating effect around the Si nanowires, since the change of temperature under 127 mW cm$^{-2}$ laser is less than $10^{-4}$ K over 100 seconds as shown by the numerical simulation by COMSOL (Fig. S3). Thus, it is reasonable to neglect both temperature and viscosity changes in the experiments.

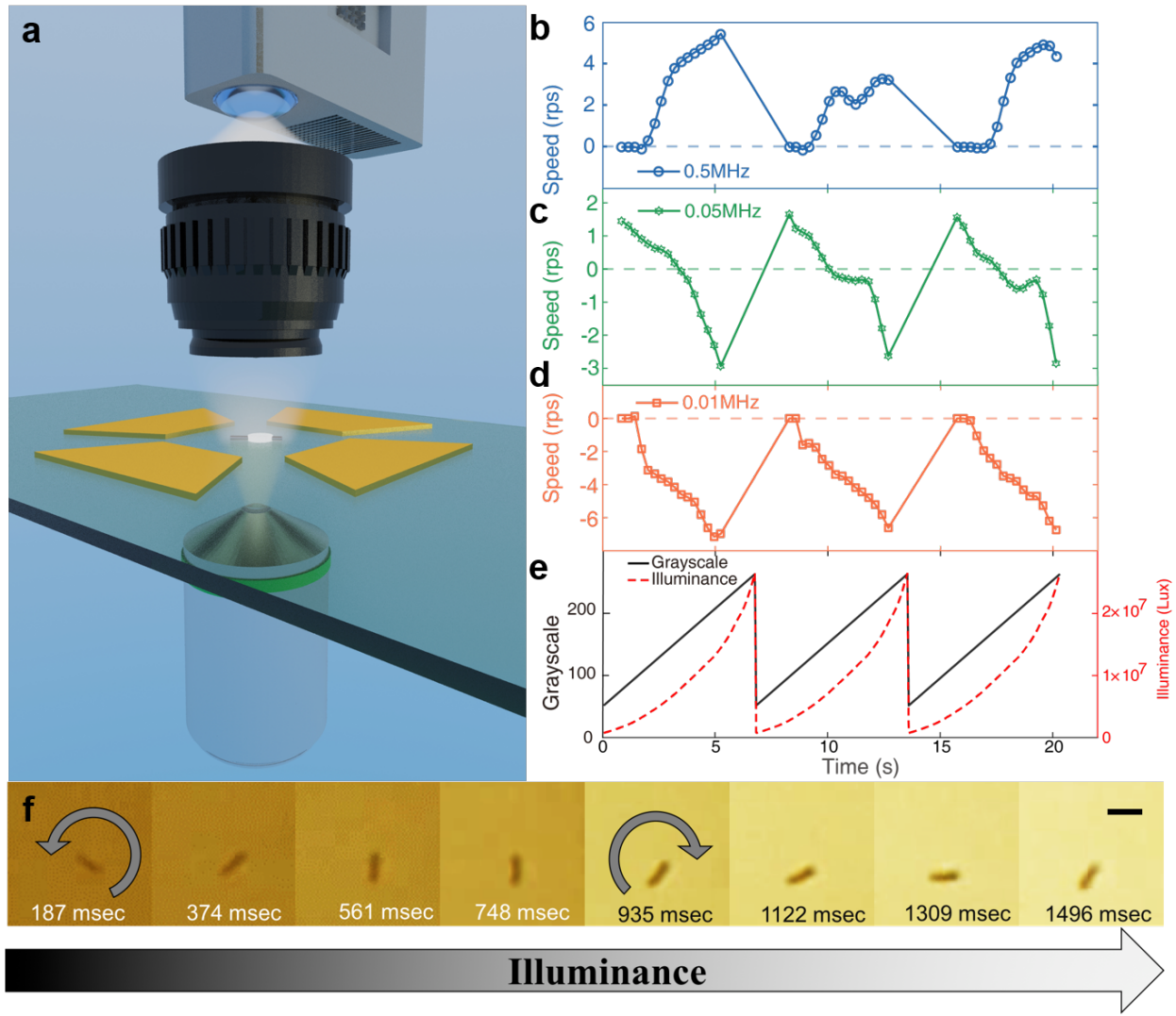

Figure 6. Computer programmable white light for reconfiguration of the silicon nanowire rotation. (a) Schematic of the white-light projection system. (b-e) Rotation speed of nanowire versus light illuminance at three different AC frequencies. (b) Acceleration in both co-field and (d) counter-field rotation, (c)

reversal of rotation; (e) corresponding illuminance as a function of pixel grayscale set in the projector. (f) Snapshots of rotation reversal with increase of illuminance (green curve in d). Scale bar, 5 μm.

**Dynamic Reconfigurable Rotation of Si Nanowires Gated by White Light.** Silicon has a broad band absorption in the visible light region. Although all the aforementioned reconfigurable rotation of Si nanowires is demonstrated by monochromatic lasers from 532 nm to 785 nm. It may not be necessary to use lasers to induce the desired opto-reconfigurable rotation of Si nanowires. Here we tested the application of white light generated by a computer controlled commercial projector to dynamically reconfigure the rotation of Si nanowires.

We customized an optical inverted microscope by equipping it with a commercial light projector (BenQ MW519) as the light source and a 6.5 mm fisheye lens (Opteka) placed above the microelectrodes for light focusing as shown in Fig. 6(a). The microelectrodes are placed in the focal plane of the fisheye lens. The commercial projector is operated with a digital light processing (DLP) system with $1280 \times 720$ pixel in resolution. We can readily design and project any colors and patterns of light to the nanowires with a resolution of $2\,\mu m \times 2\,\mu m$ per pixel. Here, we simply project a uniform monochromic rectangle, covering the entire central area of the microelectrodes and use a grayscale spectrum from 0 to 255 to adjust the illuminance, where 0 refers to a completely dark image and 255 refers to the maximum illuminance.

We increased the grayscale of the projected image from 50 to 255 at a constant rate of 30 grayscale values per second, repeated, and recorded the rotation behaviors of Si nanowires simultaneously. The illuminance as a function of grayscale is non-linear and shown in Fig. 6f. Three different rotating AC fields at 0.5 MHz, 0.05 MHz and 0.01 MHz are applied. We observed all four types of reconfigurations of rotation of Si nanowires, including acceleration, deceleration, and reverse of chirality of co-field rotation, and acceleration of counter-field rotation as shown in Fig. 6 b-d. The rotation speed versus the grayscale (illumination) of the light pattern repeated for three cycles is obtained at each AC frequency. At 0.5 MHz, the speed of the co-field rotation gradually increases with the grayscale (illuminance) (Fig. 6b, Movie S12). At 0.05 MHz, the nanowire begins with a co-field rotation in dark (grayscale 50), decreases to zero in speed, and then reverses the rotation direction as the grayscale (illuminance) further increases (Fig. 6c, Movie S11). At 0.01 MHz, the nanowire begins with a counter-field rotation in dark (grayscale 50) and increases in speed with the increase of grayscale (illuminance) (Fig. 6d, Movie S10). All types of reconfigurable rotations are repeated instantly by cycling the projected light patterns. Thus, plain white light can be used to dynamic reconfigure rotation of Si nanowires.

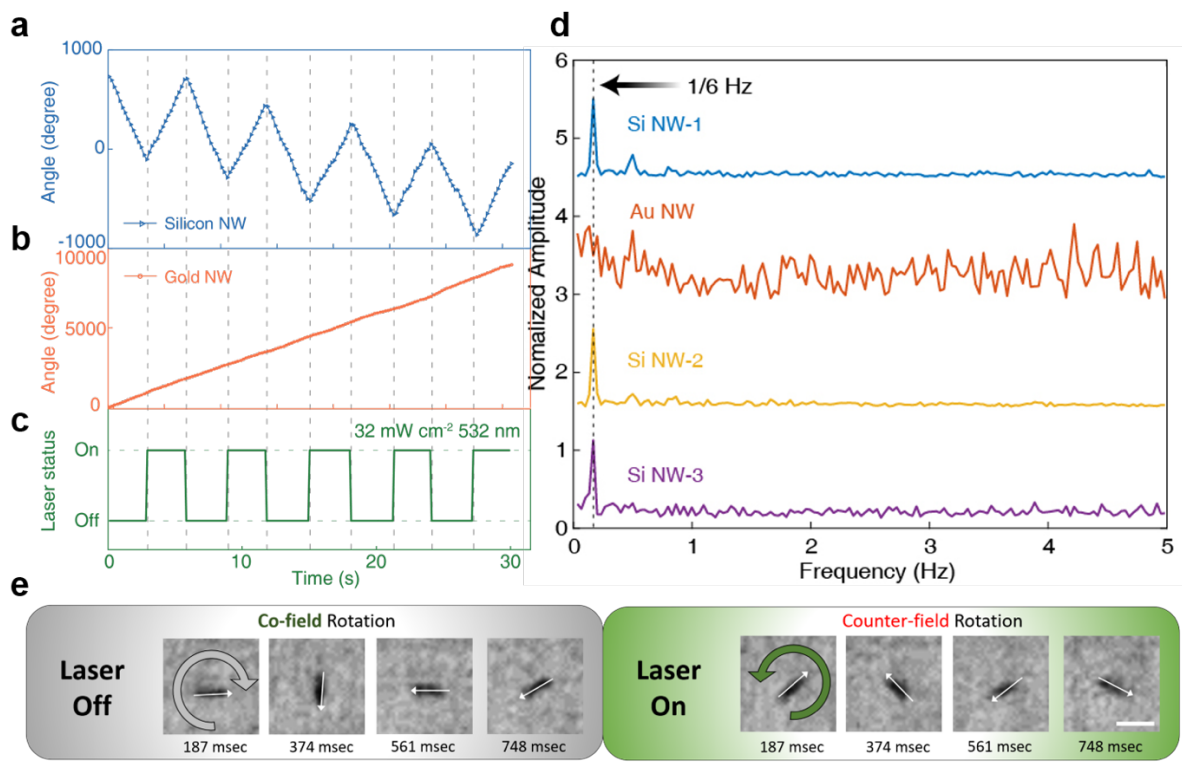

Figure 7. Different responses in rotation of semiconductor and metallic nanowire under laser stimuli in electric fields. Angle versus time of (a) silicon and (b) gold nanowire. (c) Laser of 32 $\mathrm{mW\ cm^{-2}}$ 532 nm is on and off alternatively. (d) Fourier analysis of rotation angle increment per frame. All amplitudes are normalized. The peaks at 1/6 Hz, same as that of the periodic laser stimuli, are only found in silicon nanowires. (c) Snapshots of silicon nanowire-1 showing distinct rotating behavior when laser is on or off. Scale bar, 5 μm.

**Distinguish metal and semiconductor nanowires in a mixture.** Finally, we applied this innovative opto-reconfiguration technique to distinguish metallic and semiconductor nanowires in a non-contact and non-destructive manner. Our experiments and analysis discussed above clearly show that the optical reconfigurable rotation of semiconductor Si nanowires originates from the photoconductive effect in semiconductors. While, for metallic nanoparticles, with much higher density of electrons compare to that in semiconductors, the illumination cannot change their electric conductivity. With this understanding, we demonstrated the application in distinguishing metallic and semiconductor nanowires by a plain optical stimulation in a rotating *E*-field. For a proof of concept, we mixed intrinsic silicon nanowires with gold nanowires in the same suspension in the test. It is difficult to distinguish the metallic and semiconductor nanowires with bright-field optical microscopic imaging. However, when we subject the nanowires to 5 kHz rotating AC fields and periodic illumination of 532 nm laser at 32 $\mathrm{mW\ cm^{-2}}$, distinct behavior happens for metallic and semiconductor nanowires. The semiconductor Si nanowires show periodic changes in rotation synchronized with the laser illuminance, and metallic nanowires rotate at constant speed with no response to the laser as shown in Fig. 7a, b (supplementary movie S13). For silicon nanowires, various optical responses can be observed including acceleration, deceleration, reversal rotation as well as stopping depending on the geometries and electrical properties. Though the reconfiguration of the Si nanowires can be different, and some is hard to distinguish from fluctuations caused by Brownian motion, after taking Fourier analysis of the plot of angle versus time, it is straightforward to identify silicon nanowires by the existence of a peak at the illumination frequency (1/6 Hz in Fig. 7d). While the spectrum for gold nanowire is flat through all frequencies with no obvious peaks. In this way, we can readily determine the semiconductor Si nanowires and metallic Au nanowires in the same suspension with high accuracy. Based on this working principle and with further technical development, identification of materials with less distinction could be achieved. For instance, semiconductor nanomaterials with different band gaps and absorption spectra could be distinguished by observing and analyzing their rotation behaviors in electric fields under light stimuli of a series wavelengths.

## Discussion

In the work, we discovered a new working mechanism that can readily reconfigure mechanical motions of semiconductor nanowires including acceleration, deceleration, stopping, and reversal of rotation merely by controlling the external light intensity in an electric field. This device can be considered as a mechanical analogy of field-effect transistors gated by light. We understood the new finding with both theoretical analysis and numerical simulations, and successfully demonstrated its application in distinguishing semiconductor from metal nanowires in the same suspension. With this discovery, potentially, various nanoentities, devices, and even biological cells could be equipped with mechanical responsiveness and multifold reconfigurability with functionalized semiconductor elements, changing their motion paradigms from passive to dynamic. Individually controlled micro/nanomachines amidst many could be achievable, which couple and reconfigure their operations instantly. This research could open up many opportunities for interdisciplinary fields, including reconfigurable optical devices, NEMS, nanorobots, nanomachines, communication, tunable molecule release, nanoparticle separation, and microfluidic automation.

## Methods

**Fabrication of silicon nanowires.** The silicon nanowires used in this work were fabricated via two similar methods depending on the diameter. For nanowires diameter greater than 100 nm, nanosphere lithography masked metal-assisted chemical etching (MACE) is used[38, 39]. First, all silicon wafers of various doping density from intrinsic (>5000 $\Omega\,cm$) to highly doped ($0.001\sim0.005\ \Omega\,cm$) are rinsed and sonicated in acetone, isopropanel (IPA) and DI water each for 5 minutes sequentially. After drying in nitrogen gas flow, the wafers were immersed in Poly(diallyldimethylammonium chloride) (PDDA) (2%wt), Poly(sodium 4-styrenesulfonate) (PSS) (2%wt) and Aluminum Chlorohydrate (ACH) (5%wt) each for 30 seconds that the surface can be modified with positive/negative/positive charge layers. Next, after immersing the wafer into a PS sphere suspension for 2 minutes, a monodispersed layer of negatively charged polystyrene nanospheres is formed on the wafer. Subsequently, a reactive ion etching process was carried out to slightly reduce the size of PS spheres as well as to remove the charged polyelectrolyte layers. Then a thin silver film of 50 nm in thickness was deposited on the wafer by electron-beam evaporation. Next, we sonicated the wafer in toluene to remove the PS spheres, leaving arrays of nanoholes on the silver film. The etching was conducted in the mixture of 4.7 M HF and 0.3 M $H_2O_2$. Finally, silver was removed by nitric acid and nanowires were sonicated off the wafer in DI water. For nanowires with diameter less than 100 nm, one step synthesis was adopted by directly immersing the cleaned wafer into the mixture of 4.7 M HF and 0.02 M $AgNO_3$.[41] After the etching, we treated the wafer with nitric acid to remove the silver dendrite and then sonicated nanowires off the wafer in DI water.

**Data availability.** The authors declare that the data supporting the findings of this paper are available from the corresponding author upon request.

# Acknowledgements

We are grateful for the support of National Science Foundation via the CAREER Award (Grant No. CMMI 1150767) and research grant EECS- 1710922, and Welch Foundation (Grant No. F-1734). We also thanks Dr. Chao Liu, Dr. Kwanoh Kim, Jianhe Guo Zhengtianye Wang and Daniel Teal for the helpful technical support and discussions.